# Heart Disease Detection using Quantum Computing and Partitioned Random Forest Methods

Hanif Heidari[1], Gerhard Hellstern[2], Murugappan Murugappan[3,4,5*]

1School of Mathematics and Computer Science, Damghan University, Damghan, Iran. (heidari@du.ac.ir)
2Center of Finance, Cooperative State University Baden-Württemberg, Stuttgart, Germany (Gerhard.Hellstern@dhbw-stuttgart.de)
3Intelligent Signal Processing (ISP) Research Lab, Department of Electronics and Communication Engineering, Kuwait College of Science and Technology, Block 4, Doha 13133, Kuwait (m.murugappan@kcst.edu.kw)
4Department of Electronics and Communication Engineering, School of Engineering, Vels Institute of Sciences, Technology, and Advanced Studies, Chennai 600117, India
5Center of Excellence for Unmanned Aerial Systems (CoEUAS), Universiti Malaysia Perlis, Arau 02600, Perlis, Malaysia
*Corresponding Author

**Abstract.** Heart disease morbidity and mortality rates are increasing, which has a negative impact on public health and the global economy. Early detection of heart disease reduces the incidence of heart mortality and morbidity. Recent research has utilized quantum computing methods to predict heart disease with more than 5 qubits and are computationally intensive. Despite the higher number of qubits, earlier work reports a lower accuracy in predicting heart disease, have not considered the outlier effects, and requires more computation time and memory for heart disease prediction. To overcome these limitations, we propose hybrid random forest quantum neural network (HQRF) using a few qubits (two to four) and considered the effects of outlier in the dataset. Two open-source datasets, Cleveland and Statlog, are used in this study to apply quantum networks. The proposed algorithm has been applied on two open-source datasets and utilized two different types of testing strategies such as 10-fold cross validation and 70-30 train/test ratio. We compared the performance of our proposed methodology with our earlier algorithm called hybrid quantum neural network (HQNN) proposed in the literature for heart disease prediction. HQNN and HQRF outperform in 10-fold cross validation and 70/30 train/test split ratio, respectively. The results show that HQNN requires a large training dataset while HQRF is more appropriate for both large and small training dataset. According to the experimental results, the proposed HQRF is not sensitive to the outlier data compared to HQNN. Compared to earlier works, the proposed HQRF achieved a maximum area under the curve (AUC) of 96.43% and 97.78% in predicting heart diseases using Cleveland and Statlog datasets, respectively with HQNN. The proposed HQRF is highly efficient in detecting heart disease at an early stage and will speed up clinical diagnosis.

## 1- Introduction

A change in lifestyle during the 4th Industrial Revolution has led to unhealthy conditions such as physical inactivity, obesity, diabetes, and high cholesterol that affect heart health. Over the past few years, the number of deaths caused by heart disease has increased significantly. The number of deaths caused by heart disease in 2016 was 17.7 million, and it is expected that in 2060 this number will be more than 48 million [1]. Increasing cardiac mortality rates are a global health problem. According to the WHO statistics of 2016, 82% of cardiovascular patients live in low- or middle-income countries. Additionally, cardiovascular diseases account for 50 percent of deaths in developed countries [2]. Heart disease has a significant impact on patients' quality of life [3] as well as the country's economy [4–6]. In Europe, the government spends more than 200 billion euros on patients with heart disease [5]. Moreover, it increases the risk of some other disorders such as dementia [7], pneumonia [8], cognitive dysfunction [9], and Alzheimer's disease [10]. Detecting heart disease at an early stage reduces the mortality rate associated with it [11], reduces the risk of other diseases associated with heart disease, and reduces the amount of money spent by the government on health care. In recent decades, early detection of cardiovascular diseases has become a hot topic in research. Heart disease can be categorized in five groups i.e. blood vessel disease, irregular heartbeat, congenital heart defect, disease of heart muscle and heart valve disease [12]. Coronary artery disease lies in the blood vessel disease group which is the most common heart disease in the world. Coronary heart disease is the main reason for immortality in patients with heart disease [13].

Random forests (RF) and linear methods were used simultaneously by Mohan et al. for the prediction of cardiovascular diseases [14]. They have introduced an improved version of RF for predicting early-stage coronary disease. They considered the Cleveland dataset and achieved 88.4% accuracy and 87.7% AUC. A modified deep neural network (DNN) has been developed by Khan for the prediction of cardiovascular diseases by Internet of Things (IoT) devices [15] where 98.2% accuracy was obtained. Dulhare applied particle swarm optimization (PSO) and a naive Bayes (NB) method to heart disease prediction [15]. Their results show PSO outperforms Genetic Algorithm (GA) with 87.91% accuracy for the Cleveland dataset. Waris and Koteeswaran investigated the accuracy of RF, k-nearest neighbor (KNN), NB, XGBoost, CatBoost and LightGBM for early prediction of coronary heart disease using the Cleveland dataset. They found that RF is more accurate than other methods. Maraten and Goudarzi introduced a fuzzy rule-based method for the early prediction of heart disease by considering the Cleveland dataset. They found that the Cleveland dataset did not consider some important factors such as BMI which affect heart disease. Therefore, the accuracy of the prediction will be improved by modifying the considered attributes. Zhenya and Zhang introduced a new ensemble classifier for predicting heart disease. They found AUC of 89.5% and 92.68% for the Cleveland and Statlog dataset, respectively.

Some classical machine learning algorithms such as various neural networks [16] and reinforcement learning models [17] are often limited by their computational power and computational time. A Quantum machine learning (QML) algorithm was introduced to deal with the problem in [18,19]. QML algorithms are based on quantum mechanical principles, such as superposition and entanglement. A superposition property allows an algorithm to evaluate the effects of different positions simultaneously on a system. Therefore, it may significantly reduce computation time. In essence, entanglement is the property which states particles even distant from each other are connected. As a result, QML algorithms can accelerate computations significantly and are suitable for complex or big data problems [18]. Combined with rapid advancements in communication and information technologies, QML has become a popular topic both practically and theoretically. In order to use real quantum devices, Perez-Salinas et al. [20] used a re-uploading method to overcome memory limitations. The Helstrom centroid measure was used by Sergioli et al. [21] to improve the quantum binary classification accuracy. Hellstern [22] introduced a hybrid quantum-classical neural network (HQNN) algorithm for classification problems. He found that QNN in general suffers from overfitting but is able to speed up the training, compared to a classical neural net with the same number of parameters.

Although QML algorithms are popular and play a significant role in predicting heart diseases, only a few approaches have used QML algorithms for cardiovascular disease prediction. In [19], Narain et al applied a quantum neural network to the prediction of heart disease. They considered eight features and found 98% accuracy in their gathered dataset which contains 572 records. Kumar et al. [18] investigated quantum random forest (QRF), quantum k-nearest neighbor (QKNN), quantum decision tree (QDT), and quantum Gaussian Naïve Bayes (QGNB) for detecting heart failure. They found 89% accuracy for the Cleveland dataset with 14 features. Leema et al. [23] investigated quantum particle swarm optimization as a means of predicting early heart disease. Abdulsalam et al. applied quantum support vector machine on the Cleveland dataset where AUC of 90% was obtained. Kavitha and Kaulgud used QKNN for early-stage heart disease detection [24]. They showed that data normalization and outliers removals improve the classification accuracy significantly. Alsubai et al. introduced a quantum deep learning method for early prediction of heart disease. They considered the Cleveland dataset and achieved a maximum AUC of 95% [25].

The major contributions of the paper are given below:

(a) Improved the prediction rate of heart diseases at an early stage by using quantum computing methods for either a small or large data set with limited computational complexity (in terms of computation time and memory).
(b) Developed robust quantum computing methods that can handle outlier effects in the dataset and compare their performance with existing quantum computing methods, such as HQNN.
(c) Proposed a hybrid quantum random forest (HQRF) for predicting the development of coronary heart disease in the early stages. According to the experimental results, the proposed HQRF algorithm predicts heart diseases with higher accuracy (higher AUC) than earlier methods.

As an outline of the rest of the paper, as follows: Section 2 discusses the features and datasets that are considered, and Section 3 describes how the proposed method is implemented. Detailed numerical results are presented in Section 3. The conclusion of the present work is presented in Section 4.

## 2- Materials and Methods

### 2.1 Dataset description

Features related to expressing demographic information, clinical history information, presenting symptoms, physical examination results, laboratory data and electrocardiogram (ECG) analysis results can be considered for predicting early coronary artery heart disease [26]. In this study, we examine the Cleveland and Statlog datasets from the UCI machine learning repository which are well-known datasets in the early prediction of coronary artery disease. These datasets have been used extensively to investigate the efficiency of machine learning methods in predicting heart disease early. Dulhore [27] used the Statlog dataset to investigate the effectiveness of the hybrid naïve-Bayes-PSO method in predicting heart disease. She achieved 88% accuracy by using 60 iterations in PSO. The Cleveland and Statlog datasets were used by Ayon et al. [28] to compare the accuracy of DNN, decision tree, KNN, logistic regression, NB, RF and SVM methods in predicting heart disease. Their results show that SVM with 97.36% and DNN with 98.15% accuracy are the most accurate methods for the Cleveland and Statlog datasets respectively.

According to Fitriyani et al. [29], Cleveland and Statlog datasets were analyzed using HDPM where 98.40% and 95.90% accuracy are obtained for the Cleveland and Statlog dataset respectively. In [30], Amin et al applied their feature selection method to Cleveland and Statlog data sets. They found that considering nine features by a vote method and using NB and logistic regression procedures is suitable for early heart disease prediction where 87.41% accuracy was obtained. Based on graph Lasso and Ledoit–Wolf shrinkage, Karadeniz et al. [31] introduced an ensemble method for detecting heart disease in the Statlog data. They achieved 85.5% precision for the Statlog dataset. According to Ahmad et al., considering Cleveland, Hungary, Switzerland, and Long Beach V dataset simultaneously and using RF result in 100% accuracy [32]. El-Shafiey [33] developed a hybrid PSO-random forest method for the prediction of heart disease and obtained AUC= 92% and 91% in 10-fold cross-validation for Cleveland and Statlog respectively.

### 2.1.1 Cleveland dataset

The Cleveland dataset consists of 303 data samples with 75 features [34]. There were two levels in the data that were considered for heart disease rate: individuals with a vessel diameter narrowing of less than 50% or greater than 50%. In general, the dataset has five different classes namely, normal control which is represented with a target label of 0, and labels 1 to 4 represent the four different types of heart diseases. However, the number of samples in each class of heart disease is very small and hence we combined all four classes of heart disease into one class and developed the methodology for the binary classification problem (group zero: "normal control"; group one: " coronary heart disease (CHD)").This proposal has the advantage that the proposed methodology can identify any type of heart disease as soon as the data enter into the model. Several studies have identified only 14 features as sufficient for predicting heart disease since they reduce the computational cost of the study while the accuracy of the prediction does not change significantly. In this regard, we also take into consideration the 14 features that are listed in Table 1. The data consists of 303 samples, 138 of which are healthy and 165 of which are CHD patients. A total of six samples have missing values in the data set. In order to impute missing values for each group, the median of each group was used as imputation method.

**Table 1.** The considered features for early heart disease prediction.

| Name | Feature No | Name | Feature no |
|---|---|---|---|
| Max heart rate achieved (Thalch) | 8 | Age | 1 |
| Exercise-induced angina (Exang) | 9 | Sex | 2 |
| ST depression induced by exercise relative to rest (Oldpeak) | 10 | Type of Chest pain (Cp) | 3 |
| ST segment slope (Slope) | 11 | Resting blood pressure (trestbps) | 4 |
| Major no of vessels (Ca) | 12 | Cholesterol (Chol) | 5 |
| Nuclear stress test results (Thal) | 13 | Fasting blood sugar (fbs) | 6 |
| Target (CHD/Normal) (Label) | 14 | Resting ECG test (Restecg) | 7 |

### 2.1.2 Statlog dataset

The Statlog dataset includes 270 samples with 14 features. These features are similar to those of Cleveland [29]. There are no missing values in the dataset, and 150 samples belong to group 0 (healthy patients) and

120 to group one (disease patients). Statistical properties and a comparison between Cleveland and Statlog datasets are presented in [33]. In addition, Simmons investigated the Statlog dataset's properties and concluded that it was a subset of the Cleveland dataset [13]. In contrast, researchers found different results using Cleveland and Statlog datasets, so further research should examine the relationship between statistical properties and the accuracy of machine learning methods. Some properties of the mentioned datasets are summarized in Table 2.

**Table 2:** Properties of Cleveland and Statlog datasets.

| Datasets | Number of samples | Number of features | Number of outliers data |
|---|---|---|---|
| **Cleveland** | 303 | 14 | 6 |
| **Statlog** | 270 | 14 | 3 |

## 2.2 Quantum random forest method

The field of quantum computing has become increasingly popular in recent decades. The application of quantum machine learning to practical problems has been explored by many authors. Despite considerable efforts, quantum machine learning suffers from some drawbacks that sometimes lead to questionable results. As an example, quantum computing cannot learn chaotic or random processes [35]. A quantum machine learning algorithm also faces limitations because it must deal with noise, connectivity in training networks, and parameter tuning. Additionally, quantum computation requires larger problem sizes real quantum computers, which are still in the early stages and not always available to the public [22]. Due to these limitations, quantum algorithms cannot be used for real-world problems involving big data. Compared to classical and quantum algorithms, hybrid quantum-classical models have many advantages. Hybrid classical-quantum algorithms can improve computation speed and reduce computational costs significantly [36]. Here, we propose a hybrid classical-quantum algorithm for binary classification based on random forests and quantum neural networks.

**Algorithm 1.** The hybrid quantum neural network for training

---

**Input:** *training data, numbers of qubits and layers, classical optimization algorithm with appropriate hyper-parameters*
**Output: Trained quantum neural network**
1. *Feed the data into a neural network where the dimension of the output layer is given by three times the number of qubits.*
2. *Apply the rotation gates $Ry(\phi_{i,1})Rz(\phi_{i,2})Rz(\phi_{i,3})$ on each qubit. Here, the three angles are determined by the output of the classical layer. i is the index of the qubit.*
3. *Entangle the qubits by CNOT gates.*
4. *Apply rotation and CNOT gates on each layer again but where the three angles are now variational parameters.*
5. *Measure each qubit and feed the results into a second neural net with two neurons.*
6. *Apply the steps 3-5 for each input sample and build an appropriate objective function (binary cross entropy)*
7. *Apply a classical optimization algorithm for finding the unknown weights of the neural networks and the variational parameters.*
8. *Evaluate the Area Under Curve (AUC) of the method by applying the model to training and test data*

---

*2.2.1 Random Forest method*

Random forest methods are used to classify and predict data using ensemble methods. The RF method is generally based on decision trees [33]. Consider a binary classification problem with $k$ features. The algorithm partitions the features into $m$ subsets, and the decision tree method is applied to each subset. The parameter $m$ and the tree depth are hyper-parameters by the algorithm and must be specified.

*2.2.2 Quantum neural network*

Quantum neural networks (QNN) are modifications of artificial neural networks whose inputs and training procedures are based on quantum mechanics principles. Quantum networks are composed of qubits as input, and their dynamics follow the rules of superposition and quantum entanglement [20,22]. For N-dimensional problems, the number of qubits needs to be at least $log_2 N$ if amplitude encoding is considered [20]. This is a restriction to quantum computing. It is, fortunately, possible to solve a general optimization problem with an arbitrary dimension from just one qubit coupled with a classical optimization problem. As a result of this fact, Pérez-Salinas et al. [20] introduced data-reuploading as a solution to multidimensional classification problems that rely on just one quantum qubit, utilizing a variational quantum formulation. Hellstern [22] recently introduced a hybrid quantum neural network (HQNN) based on re-uploading data and reducing features via a classical neural network in front of the quantum network. The procedure is outlined in Algorithm 1.

**Algorithm 2.** The hybrid quantum random forest method.

---

**Input:** *training data, numbers of qubits (nq) and layers, number of features (m).*
**Output: Trained quantum neural network**
1. Partition the feature set into $D = ceil[m/(3nq)]$ where each partition has at most $3*nq$ elements.
2. For $i = 1 \dots D$ do the Steps
3. Apply the rotation gates $Ry(\phi_{i,1})Rz(\phi_{i,2})Rz(\phi_{i,3})$ on each qubit. Here, the three angles are determined by the output of the classical layer. i is the index of the qubit.
4. Entangle the qubits by $CNOT$ gates.
5. Apply rotation and $CNOT$ gates on each layer again but where the three angles are now variational parameters.
6. Measure each qubit and feed the results into a second neural net with two neurons.
7. Apply the steps 3-6 for each input sample and build an appropriate objective function (binary cross entropy)
8. Apply a classical optimization algorithm for finding the unknown weights of the neural networks and the variational parameters.
9. Aggregate the results over the D "trees" and evaluate the AUC for training data. When the parameters are fixed apply the model to test data and calculate the AUC.

---

### 2.2.3 Hybrid quantum neural random forest

The RF algorithm is a popular machine-learning method used to solve regression and classification problems. Even though RF is widely used, it suffers from overfitting and computational complexity. Additionally, QNNs are inefficient in solving high-dimensional problems since they require a high amount of memory and computational power on noisy-intermediate computers. QNN is also limited by the number and capacity of pure quantum computers, which restricts its application. We therefore propose a hybrid quantum neural network and random forest (HQRF) in this subsection that overcomes the inefficiencies of QNN and RF.

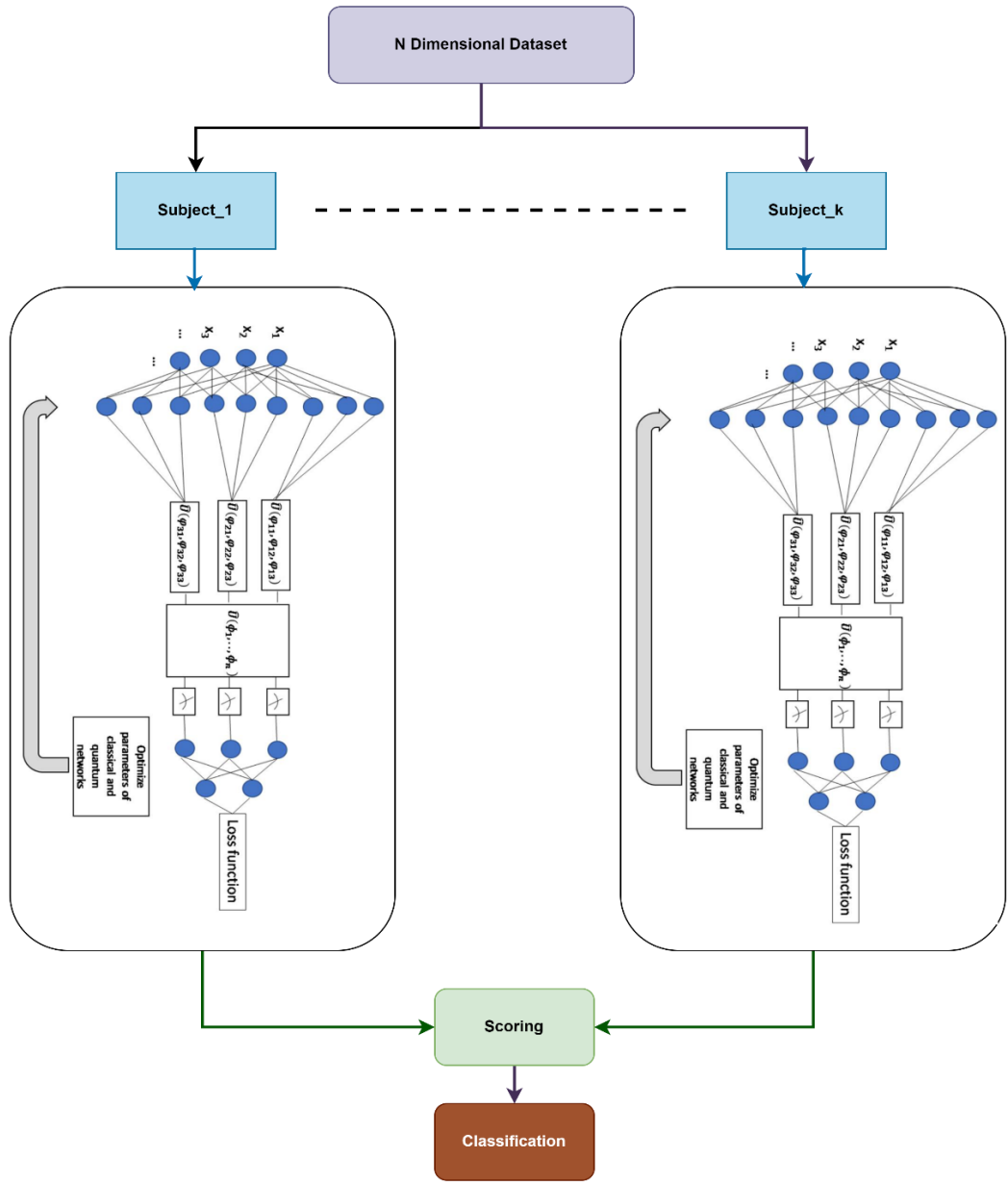

**Figure 1.** The diagram of HQRF.

A random forest algorithm is applied to the problem in the first step. This step involves splitting up a high-dimensional problem into independent low-dimensional problems (subproblems). The hybrid QNN is then applied to each subproblem. By using distributed computers to solve the subproblems in parallel, run times can be reduced significantly (Figure 1). As a final step, the same traditional random forest is used to determine the predicted class based on the score of each class. A description of the procedure can be found in Algorithm 2. In HQRF, the number of trees and tree depth are not considered as input parameters since tree depth does not matter and the number of trees is dependent on the number of qubits present. Using the QNN algorithm [22], we can consider at most 3* number_of_qubits variables in each subproblem. Due to this, the proposed algorithm requires fewer input parameters than RF and improves the overfitting problem. In this article, we partition the 14 features into four subsets, each with five, five, three, and one feature, where the first 13 elements are inputs and the last one is outputs. The algorithm was run five times independently where the features were partitioned randomly each time. The results are summarized in the next section.

## 3 Numerical results and Discussion

In this work, we have compared the performance of our proposed HQRF with our earlier algorithm (HQNN) for a different number of qubits and layers. Since, HQNN has been most widely used for heart disease prediction in the literature [22]. There are various parameters that can be used to assess the efficiency of machine learning algorithms. Accuracy (Acc), sensitivity (Sen), specificity (Spe), positive predictive value (PPV), F1-score, and area under the curve (AUC) are some of the best-known metrics used in evaluating algorithms [29,33,37]. In this case, we have N elements that need to be classified in a binary manner. A class is labeled with 0 or 1 based on the outcome, where 0 represents no disease (healthy control) and 1 represents coronary heart disease (CHD). Table 3 describes the notations used to define the evaluation measure mentioned above. The accuracy, sensitivity, specificity, positive predictive value, and F1 score formulas are defined in Table 4.

**Table 3.** Notations and descriptions of true positive (TP), false positive (FP), false negative (FN), and true negative (TN) in the heart disease datasets.

| Description | Notation |
|---|---|
| The number of true positive identified members i.e., the number of true predicted elements in class 1. | TP |
| The number of true negative identified members i.e., the number of true predicted elements in class 0. | TN |
| The number of false positive identified members i.e., the number of predicted elements in class 1 that are actually in class 0. | FP |
| The number of false negative identified members i.e., the number of predicted elements in class 0 that are actually in class 1. | FN |

**Table 4.** The considered efficiency measures for comparing the classification algorithms.

| Performance Measures | Formula | |
|---|---|---|
| **Accuracy (Acc)** | $Acc = \frac{TP+TN}{TP+TN+FP+FN}$ | (1) |
| **Sensitivity (Sens)** | $Sens = \frac{TP}{TP+FN}$ | (2) |
| **Specificity (Spec)** | $Spec = \frac{TN}{TN+FP}$ | (3) |
| **Positive Predictive Value (PPV)** | $PPV = \frac{TP}{TP+FP}$ | (4) |
| **F1-score (F1)** | $F1 = 2\frac{PPV'Sens}{PPV+Sens}$ | (5) |

Based on the receiver operating characteristic (ROC) curve, the AUC parameter is also calculated. With $x$ being $Sens$ and $y$ being $1 - Spec$, we construct the ROC curve by using different thresholds for points $(x, y)$. The AUC is often the most desired efficiency score among the different efficiency scores since it is both consistent and discriminating [38]. Smoot et al. [39] measured AUC to compare the accuracy of different breast cancer diagnostic methods. AUC was used by Huang et al. [40] to compare the efficiency of landslide susceptibility prediction using machine learning methods; Cai et al. [41] ranked the relationship between biomarkers and event time in cardiovascular disease by AUC; Rosendael et al. [42] studied the risk of major cardiovascular events by coronary stenosis and plaque composition based on the value of AUC. As a result, we are mainly interested in comparing quantum computing algorithms proposed in earlier works based on the AUC performance measure. In order to compare the efficiency of the proposed algorithm with the existing algorithm, two distinct procedures are used, namely 10-fold cross-validation and 70/30 train-test ratio.

**Table 5.** A comparison of AUC between different methods for predicting heart disease in the Cleveland dataset using 10-fold cross-validation

| Number of Qubits=4 | | | | Number of Qubits=3 | | | | Number of Qubits=2 | | | |
| --- | --- | --- | --- | --- | --- | --- | --- | --- | --- | --- | --- |
| L=4 | L=3 | L=2 | L=1 | L=4 | L=3 | L=2 | L=1 | L=4 | L=3 | L=2 | L=1 |
| HQNN | | | | | | | | | | | |
| 95.59 | 94.5 | 94.2 | 92.03 | 93.72 | **96.43** | 92.59 | 93.05 | 88.47 | 92.61 | 91.78 | 91.73 |
| HQRF | | | | | | | | | | | |
| 88.74 | 90.36 | 88.91 | 90.37 | 89.34 | 90.27 | **91.14** | 90.15 | 88.1 | 89.34 | 90.12 | 88 |

**Table 6.** The area under the curve (AUC) for HQNN and HQRF with 70-30 (train/test) split ratio in Cleveland using different numbers of qubits and layers.

| Number of Qubits=4 | | | | Number of Qubits=3 | | | | Number of Qubits=2 | | | |
| --- | --- | --- | --- | --- | --- | --- | --- | --- | --- | --- | --- |
| L=4 | L=3 | L=2 | L=1 | L=4 | L=3 | L=2 | L=1 | L=4 | L=3 | L=2 | L=1 |
| HQNN | | | | | | | | | | | |
| 95.92/ 90.67 | 94.50/ 89.06 | 94.77/ 90.18 | **94.89/ 93.78** | 96.98/ 92.75 | 94.92/ 90.13 | 93.98/ 91.88 | 94.26/ 93.34 | 93.01/ 92.22 | 94.13/ 90.76 | 93.45/ 88.09 | 93.30/ 93.24 |
| HQRF | | | | | | | | | | | |
| 95.85/ 92.8 | 94.71/ 92.85 | 91.93/ 93.29 | **89.59/ 94.36** | 92.23/ 91.44 | 92.31/ 92.12 | 90.64/ 93.92 | 85.44/ 85.47 | 85.02/ 85.64 | 85.79/ 89.06 | 93.61/ 81.58 | 86.69/ 94.31 |

## 3.1 Cleveland dataset

The HQNN method [22] is used to predict heart disease. The results of all numerical tests are validated and compared between 10 independent runs just to ensure that they are robust. Table 5 represents the results of 10-fold cross-validation with a different number of qubits and layers (L). An HQNN with 3 layers and 3 qubits achieved a maximum AUC of 96.43% compared to other combinations and our proposed HQRF. Based on a 70/30 split ratio of train to test, we evaluated the AUCs of the train and test sets and results are reported in Table 6. The AUC of the test set is used to compare the efficiency of methods since the effectiveness of methods is typically determined by their ability to predict correct labels for new elements. It is evident from Table 6 that HQRF with only 2 qubits and 1 layer is superior to other methods, with a 94.36% AUC for the test set. We see that in some parts of Table 6, the train set is less accurate than the test set. In conclusion, HQRF can detect and ignore outlier data during the execution of the classification method. A comparison between the accuracy of the test set in the 10-fold cross (Table 5) validation and the 70/30 split ratio of train to test method (Table 6) indicates HQRF outperforms in the 70/30 split ratio of train to test method while HQNN outperforms in 10-fold cross-validation. This concludes HQNN requires larger training dataset while HQRF is more appropriate for small dataset.

In Figure 2, we show the ROC plot for the most favorable parameter combination (three qubits and three layers) of all ten folds used in cross-validation for the Cleveland dataset using HQNN. In some recent works, researchers used feature selection methods for improving the heart disease prediction accuracy of their suggested methods. Fitriyani et al. [29] introduced HDPM method where information gain (IG) feature selection method was used for improving the accuracy of the classification. El-Shafiey et al. [33] used a combination of RF and genetic algorithm for feature selection in the Cleveland and Statlog dataset. The two-tier CE [43] and CE-Ensemble [44] methods used a PSO-based and relief feature selection algorithm respectively. Although the feature selection methods lead to improvement of classification accuracy, it requires large computational efforts which is time consuming. Quantum computing algorithms gain high speed computation which overcomes the mentioned inefficiency. Figure 3 compares the accuracy of HQNN, HQRF, with the recent mentioned classification methods. Area Under Curve achieved by our two methods (HQNN and HQRF) is higher than other classification methods except for HDPM which scored 100%. HDPM employs a train/test ratio of 90/10 and includes feature selection, outlier detection, and data balancing/augmentation. Next, the classification task is performed using the XGBoost method. In a recent study, Bentéjac et al. compared the XGBoost method with the RF method on the Cleveland dataset [45]. Results show that the gradient boosting method is more accurate than others with 83.78% accuracy. Therefore, HDPM benefits from preprocessing data that is not computationally efficient. Based on our findings, we conclude that the mentioned quantum methods outperform others that do not require any preprocessing.

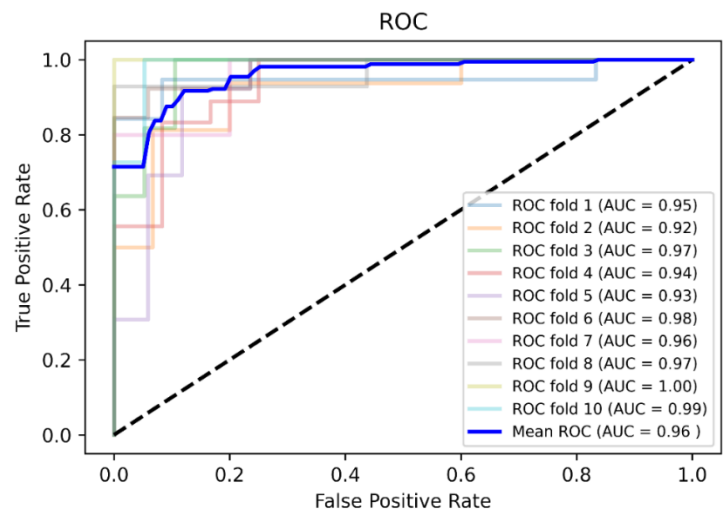

**Figure 2.** The ROC plots for each of the 10 folds using HQNN with 3 layers and 3 qubits and 10-fold cross-validation. Further the AUC value for each fold and the mean value is shown.

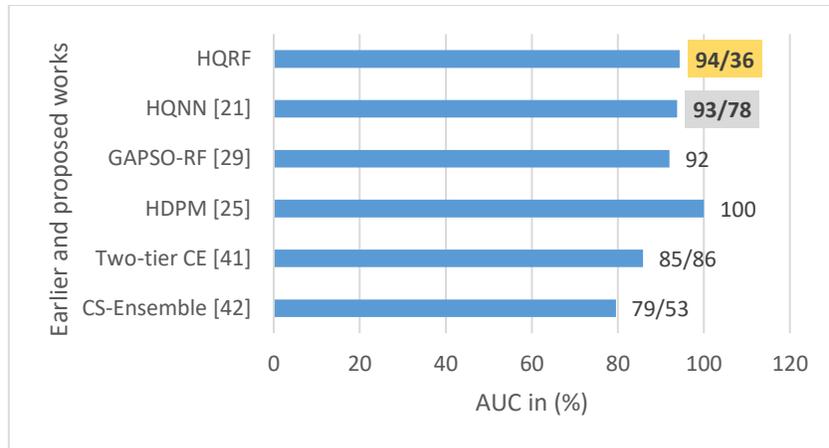

**Figure 3.** Comparison of AUC obtained in this paper with previous results in the literature.

### 3.2 Statlog dataset

Figure 4 shows the area under curve (AUC) of HQNN and HQRF on Statlog dataset with 10-fold cross-validation. The maximum mean AUC of 97.78% is achieved with HQNN (3 qubits) and 91.64% with HQRF (2 qubits). The results for 70-30 train/test split ratio with different numbers of qubits and layers are given in Table 8. As a result of the 70/30 train/test split ratio procedure, the HQRF method with three qubits and one layer performs the best. It is appropriate to use HQRF for a 70/30 train/test split ratio, and HQNN for a 10-fold cross-validation. Thus, HQNN is a better method for large datasets, whereas HQRF is better for small datasets, where both methods should be able to handle a small train set size.

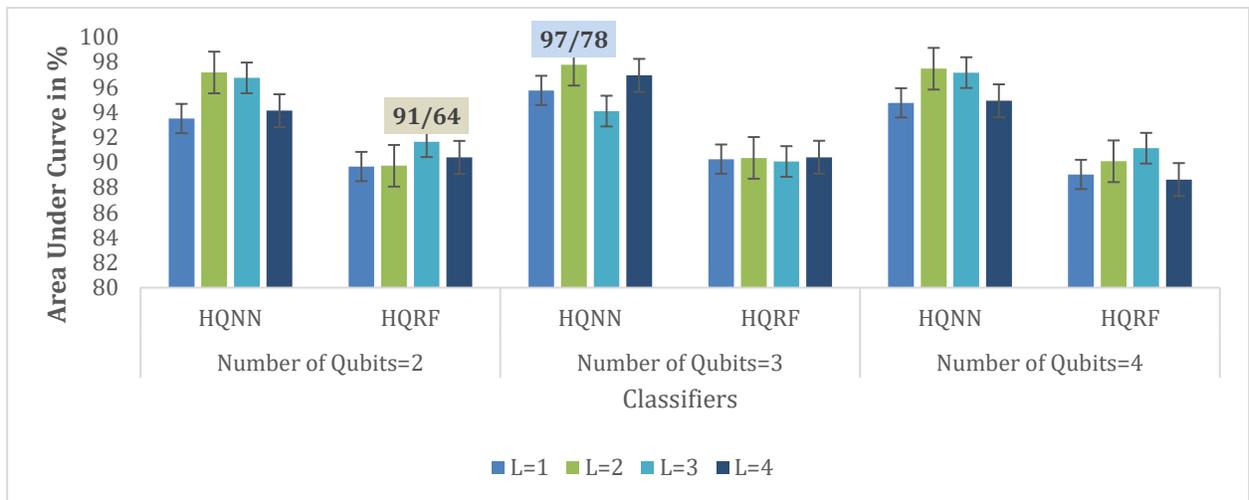

**Figure 4.** The area under the curve (AUC) for HQNN and HQRF with 10-fold cross-validation for the Statlog dataset using different numbers of qubits and layers.

**Table 7:** The area under the curve (AUC) for HQNN and HQRF with 70-30 train/test split ratio in Statlog dataset using different numbers of qubits and layers.

| Number of Qubits=4 | | | | Number of Qubits=3 | | | | Number of Qubits=2 | | | |
|---|---|---|---|---|---|---|---|---|---|---|---|
| L=4 | L=3 | L=2 | L=1 | L=4 | L=3 | L=2 | L=1 | L=4 | L=3 | L=2 | L=1 |
| HQNN | | | | | | | | | | | |
| 99.17/ 82.65 | 99.21/ 82.96 | 98.76/ 88.14 | 99.22/ 84.87 | 98.15/ 79.16 | 98.40/ 82.83 | 99.00/ 76.29 | 98.83/ 80.61 | 98.83/ 76.72 | 98.43/ 79.93 | 99.05/ 84.19 | 97.97/ 88.33 |
| HQRF | | | | | | | | | | | |
| 96.55/ 86.17 | 95.07/ 88.76 | 95.47/ 85.67 | 89.25/ 89.93 | 95.52/ 86.23 | 92.75/ 81.97 | 91.56/ 81.72 | **87.60/ 90.52** | 86.99/ 76.29 | 88.21/ 74.62 | 85.15/ 80.21 | 86.97/ 76.35 |

Table 8 is similar to Table 5, which shows that the train set's accuracy is less than the test set's accuracy at the optimal point. The results confirm that HQRF can detect outliers by considering data and ignore them when constructing classification by considering procedures. In Figure 5 we show the ROC plot for the most favorable parameter combination (three qubits and two layers) of all ten folds used in cross-validation for the Statlog dataset. The mean of AUC of 98% confirms the accuracy and robustness of HQNN for early prediction coronary artery disease.

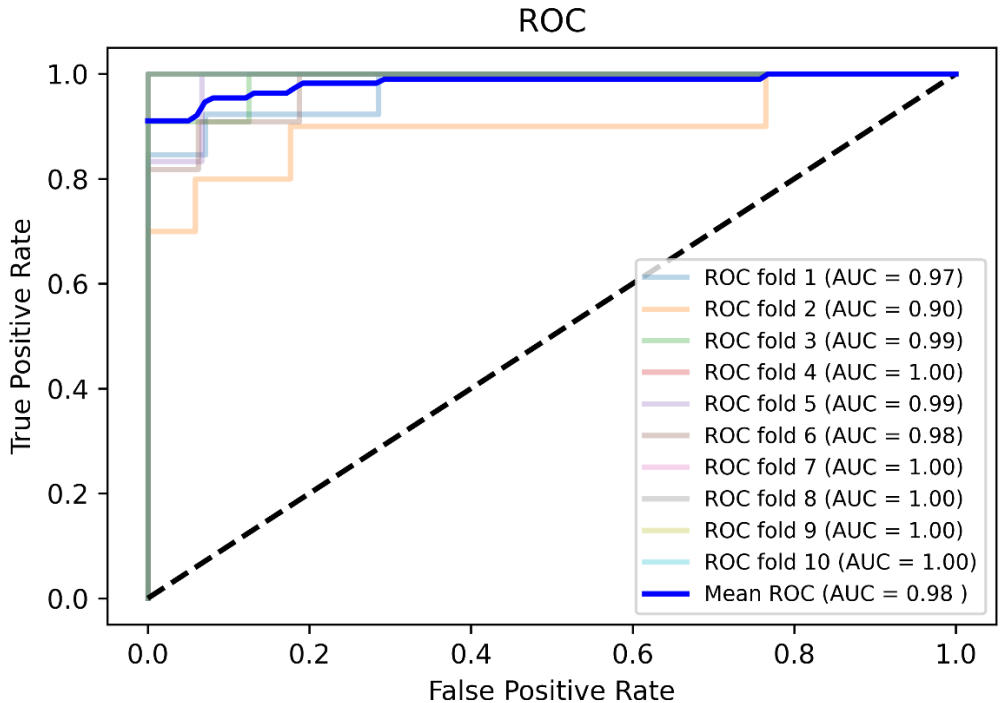

**Figure 5.** The ROC plot for the classification Statlog dataset where 2 layers and 3 qubits with 10-fold cross-validation are considered.

A comparison of the HQNN and HQRF accuracy with the recently introduced quantum computing based heart disease classification methods is presented in Table 8. Due to the reason discussed earlier, HQNN outperforms other methods except HDPM. A comparison of Table 7 with Figure 4 reveals that HQRF is not efficient for the Statlog dataset. To investigate the theoretical reasons behind this, we examine some properties of datasets as shown in Table 2. Two major differences are evident in this study: the number of samples and the number of outliers. HQNN performed better in 10-fold cross-validation cases, which indicates that a large enough training dataset is required. Furthermore, the Statlog dataset has half the number of outliers as the Cleveland dataset, resulting in more accurate results in HQRF. Therefore, we conclude that HQNN is highly sensitive to outliers while HQRF is robust under noise. Simmons also examined the Statlog dataset and concluded that it was a subset of the Cleveland dataset [13]. Therefore, HQRF outperforms in large datasets even if outliers are present. Due to its ability to partition data dimensions into smaller sets, HQRF can also be used to solve high-dimensional problems. The number of qubits in quantum computing is a critical item, so HQRF uses smaller numbers of qubits during each step of the simulation, which makes data with small dimensions better suited to quantum simulation.

**Table 8.** A comparison of AUC between different methods for predicting heart disease in Statlog dataset.

| Statlog Dataset | Two-tier CE [43] | CS-Ensemble [44] | GAPSO-RF [33] | HDPM [29] | HQNN [22] | **HQRF (Proposed)** |
|---|---|---|---|---|---|---|
| **AUC (in %)** | 93.42 | 87.99 | 92 | 100 | 97.78 | **91.64** |

## 4 Conclusion

Classification using RF is a widely used technique in practical applications. Tree depth and number of trees are the two parameters used by RF to determine its output. The RF technique fails if the tree depth and number of trees are selected incorrectly, resulting in overfitting. Additionally, QNN is time-consuming and should only be used for low-dimensional problems because of its high computational cost. To overcome these difficulties, this paper proposes HQRF and uses it to predict heart diseases at an early stage. The Cleveland and Statlog heart disease datasets are used to evaluate HQNN and HQRF. HQNN and HQRF compared with some recently introduced classification methods. As a result, the studied methods use different feature selection methods which require high computational efforts and are time consuming while HQNN and HQRF are high speed algorithms since the nature of quantum computing. The numerical results show that HQRF is more appropriate for small datasets while HQNN is better suited for larger datasets. Further, HQNN is highly sensitive to outliers, whereas HQRF has a very low sensitivity to outliers. In future work, we suggest combining an outlier detection method with HQNN to improve its efficiency.